\documentclass[aip,amsmath,amssymb,jcp, reprint]{revtex4-1} 

\usepackage{graphicx}
\usepackage{gensymb}
\usepackage[aboveskip=0pt,belowskip=-5pt]{caption}
\usepackage{subfig}
\usepackage{dcolumn}
\usepackage{bm}
\usepackage[version=4]{mhchem}
\usepackage{braket}
\usepackage[export]{adjustbox}
\usepackage[section]{placeins}
\usepackage{multirow}
\usepackage[utf8]{inputenc}
\usepackage[T1]{fontenc}
\usepackage{threeparttable}
\graphicspath{{./}}
\usepackage{cleveref}
\usepackage{gensymb}
\usepackage{diagbox}
\usepackage[dvipsnames]{xcolor}
\usepackage{simplewick}

\begin{document}
\title[]{A Stochastic Approach to Unitary Coupled Cluster}

\author{Maria-Andreea Filip}
 \email{maf63@cam.ac.uk}
\author{Alex J W Thom}
 \email{ajwt3@cam.ac.uk}

\affiliation{ 
Department of Chemistry, University of Cambridge, Cambridge, UK
}%
\begin{abstract}
Unitary coupled cluster (UCC), originally developed as a variational alternative to the popular traditional coupled cluster method, has seen a resurgence as a functional form for use on quantum computers. However, the number of excitors present in the \textit{ansatz} often presents a barrier to implementation on quantum computers. Given the natural sparsity of wavefunctions obtained from Quantum Monte Carlo methods, we consider here a stochastic solution to the UCC problem. Using the Coupled Cluster Monte Carlo framework, we develop cluster selection schemes that capture the structure of the UCC wavefunction, as well as its Trotterized approximation, and use these to solve the corresponding projected equations. Due to the fast convergence of the equations with order in the cluster expansion, this approach scales polynomially with the size of the system. Unlike traditional UCC implementations, our approach naturally produces a non-variational estimator for the energy in the form of the projected energy. For UCCSD in small systems, we find this agrees well with the expectation value of the energy and, in the case of two electrons, with full configuration interaction results. For the larger N$_2$ system, the two estimators diverge, with the projected energy approaching the coupled cluster result, while the expectation value is close to results from traditional UCCSD. 
\end{abstract}
\maketitle

\section{Introduction}
Coupled cluster\cite{Cizek1966,Cizek1969} (CC) theory has long been one of the
most popular \textit{ab initio} methods in quantum chemistry, due to its balance
of high-accuracy, size-consistency, polynomial scaling and systematic
improvability, with its truncation at single and double excitations with
perturbative triples\cite{Raghavachari1989} (CCSD(T)) being considered the
industry "gold-standard". However, to obtain these appealing features, one has
to sacrifice the variationality of the method. Unlike energies obtained from methods 
such as Hartree--Fock (HF) theory, configuration interaction (CI) or even
density functional theory (DFT), coupled cluster projected energies do not obey the 
variational principle, so there is no guarantee that they will be an upper 
bound on the ground state energy. At times, this allows for 
catastrophic behaviour, such as the well known collapse of the CCSD(T) energy 
in the case of strongly correlated systems, such as N$_2$.\cite{Chan2004} 
\newline\newline
A variational formulation of coupled cluster is therefore a tempting 
proposition and many attempts have been made to develop such formulations,\cite{Szalay1995} 
leading to methods such as variational coupled cluster,\cite{Bartlett1988, VanVoorhis2000} 
extended coupled cluster\cite{Arponen1983} and unitary coupled cluster.\cite{Kutzelnigg1982, Kutzelnigg1983, Kutzelnigg1984, Tanaka1984, Hoffmann1987, Hoffmann1988, Bartlett1989} 
However, these often involve non-terminating series for the energy, which 
cannot necessarily be truncated intuitively. Computational scaling is also
increased, becoming exponential for methods like variational and unitary 
coupled cluster.
\newline\newline
While these various issues have stopped alternative CC methods from
becoming widely adopted in the quantum chemistry community, recently there has
been renewed interest in the unitary coupled cluster formalism, due to its
applicability as a wavefunction \textit{ansatz} for quantum computers, which
are emerging as a promising tool for computational chemistry. The qubit model
of computation naturally encodes the exponentially sized Hilbert space of a 
molecule in a linear number of qubits. By mapping each qubit to a spin-orbital 
and appropriately entangling them, one can encode a FCI wavefunction in a
number of qubits equal to the number of spin-orbitals in the basis, rather than 
requiring the storage of $\begin{pmatrix} N_\mathrm{spinorb} \\ N_\mathrm{elec} 
\end{pmatrix}$ determinant coefficients. In principle, on an ideal quantum
computer, one can obtain a wavefunction with good overlap with the true ground 
state wavefunction using adiabatic state preparation\cite{Farhi2001} and trotterized time evolution\cite{Aspuru-Guzik2005, Whitfield2011}, where a guess 
wavefunction is propagated to the ground state, in an approach reminiscent of 
DMC\cite{Anderson1975,Ceperley1986} and FCIQMC\cite{Booth2009}. The true ground 
state energy can then be computed using Quantum 
Phase Estimation\cite{Kitaev1995, Abrams1999} (QPE). However, current quantum 
computers, known as Noisy Intermediate-Scale Quantum (NISQ) machines, are 
limited in both number of qubits and qubit decoherence time, as well as being 
subject to significant noise in the results, making the 
application of such a resource--intensive algorithm infeasible. However, 
alternatives tailored to such machines have been developed. One such algorithm 
is the Variational Quantum Eigensolver\cite{Peruzzo2014, McClean2016} 
(VQE) where an \textit{ansatz}-based wavefunction is prepared on a quantum 
computer, its energy is measured and a classical optimisation algorithm is 
used to minimise the energy and get optimal wavefunction coefficients. As all 
operations available to a quantum computer must be representable by a unitary 
transformation, UCC has resurfaced as an appropriate \textit{ansatz} for this 
algorithm.\cite{McClean2016, Cao2019, Romero2019}
\newline\newline
The number of excitation operators in the expansion can be a limiting factor 
in the use of UCC \textit{ansatze}, as it increases the required quantum 
circuit depth. Therefore, it would be useful to develop a method to pre-select
the most important cluster operators before devising the circuit.
Quantum Monte Carlo (QMC) methods, such as FCIQMC\cite{Booth2009} and CCMC\cite{Thom2010}
produce naturally sparse representations of the wavefunction, as only 
significant contributors are meaningfully sampled by the propagation 
scheme. Therefore, they seem ideal candidates for the screening of cluster 
amplitudes. This idea has been suggested before as a means to only include the 
most important triple and quadruple excitations in a conventional CC 
calculation, with good results.\cite{Deustua2017, Deustua2018, Deustua2019} We would be interested in 
assessing its applicability in screening UCC amplitudes as well, and will pursue this in further work. 
\newline\newline
In this paper, we set out to detail the algorithmic steps involved in the 
implementation of a projective variant of UCC, primarily in a stochastic 
framework, as well as discussing the results obtained from this approach on a 
series of test systems. In the following section, we review theoretical aspects of
coupled cluster theory and its unitary formalism. We then revisit the CCMC 
algorithm in section III and discuss its expansion to UCC in section IV. 
Section V presents a series of benchmark results and Section VI brings 
together our conclusions.

\section{Coupled Cluster Theory}
\subsection{Traditional Coupled Cluster}
\label{sec:ccmc}
In traditional coupled cluster, the wavefunction is given by an exponential 
\textit{ansatz }
\begin{equation}
\Psi_{\mathrm{CC}} = e^{\hat T}\ket{D_{0}},
\end{equation}
where $\ket{D_{0}}$ is 
the Hartree--Fock reference wavefunction and the cluster operator 
\begin{equation}
\hat T = \sum_i \hat T_i
\end{equation}
where operators $\hat T_i$ generate determinants of excitation level 
$i$, \textit{i.e.}
\begin{equation}
\hat T_1 = \sum_{i,a} t_i^a \hat a^\dagger \hat i
\end{equation}
{\color{black}\begin{equation}
\hat T_2 = \frac{1}{4}\sum_{i,j,a,b} t_{ij}^{ab} \hat a^\dagger \hat b^\dagger \hat i \hat j
\end{equation}}
and so on. In this context, $\hat p^\dagger$ and $\hat p$ represent creation 
and annihilation operators for orbital $p$ respectively. In the summations, 
$i,j,...$ range over the occupied orbitals in the reference and $a,b,...$ range over 
the virtual orbitals. The CC wavefunction is equivalent to the FCI wavefunction 
if all possible excitors are included. However, one can truncate $\hat T$ to a 
certain excitation level, giving a progression of increasingly accurate 
methods: CCSD
($\hat T = \hat T_1 + \hat T_2$), CCSDT ($\hat T = \hat T_1 + \hat T_2 + \hat T_3$), 
CCSDTQ ($\hat T = \hat T_1 + \hat T_2 + \hat T_3 + \hat T_4$) and so on. 
\newline\newline
In order to obtain $t_{\textbf i}$, the Schrödinger equation is projected 
onto each of the determinants $\ket{D_{\textbf i}}$,
leading to a series of coupled cluster equations to be solved:
\begin{equation}
\braket{D_{\textbf i}|\hat H - E|\Psi_{\mathrm{CC}}} = 0,
\end{equation} 
where $E$ is the energy of $\Psi_{\mathrm{CC}}$. One can instead use a 
similarity transformed Hamiltonian $\bar H = e^{-\hat T} \hat H e^{\hat T}$, 
giving an equivalent set of equations:
\begin{equation}
\braket{D_{\textbf i}|\bar H - E|D_0} = 0,
\end{equation} 
The Campbell--Baker--Hausdorff (CBH)\cite{Campbell1897, Baker1905, Hausdorff1906} 
expansion of the similarity transformed Hamiltonian
\begin{align}
\begin{split}
\bar H = &\hat H + [\hat H, \hat T] + \frac{1}{2} [[\hat H, \hat T],\hat T] + \frac{1}{3!}
[[[\hat H, \hat T], \hat T], \hat T] +\\
&+ \frac{1}{4!} [[[[\hat H, \hat T], \hat T], \hat T], \hat T]
\end{split}
\label{eq:cccbh}
\end{align}
terminates at fourth order, guaranteeing a finite number of terms in the 
projected CC equations. The time required to computationally solve these 
equations scales as $\mathcal O(N^{2n+2})$ where $N$ is the size of the system 
and $n$ is the truncation level. Therefore, CCSD scales as $\mathcal O(N^6)$,
CCSDT as $\mathcal O(N^8)$ and so on.
\newline\newline
Having solved these projected equations, one typically uses the projected energy
\begin{equation}
E_\mathrm{proj} = \braket{D_0|\hat H|\Psi_\mathrm{CC}}
\end{equation}
as an estimator for the energy of the wavefunction. Where coupled cluster is able to capture the true wavefunction, this should be equal to the expectation value of the energy,
\begin{equation}
\braket{E}_ \mathrm{CC} = \frac{\braket{\Psi_\mathrm{CC}|\hat H| \Psi_\mathrm{CC}}}{\braket{\Psi_\mathrm{CC}| \Psi_\mathrm{CC}}}.
\end{equation}
However, where coupled cluster is not exact, this equality is not guaranteed.

\subsection{Unitary Coupled Cluster}

Consider the anti-Hermitian operator
\begin{equation}
\hat \tau = \hat T - \hat T^\dagger,
\end{equation}
where as before $\hat T$ can be truncated in excitation level. $e^{\hat \tau}$ 
is then a unitary operator and can be used to define a UCC wavefunction
\begin{equation}
\ket{\Psi_\mathrm{UCC}} = e^{\hat \tau}\ket{D_0}
\end{equation}
The expectation value of the energy is then given by
\begin{align}
\begin{split}
\braket{E}_\mathrm{UCC} &= \frac{\braket{\Psi_\mathrm{UCC}|\hat H| \Psi_\mathrm{UCC}}}{\braket{\Psi_\mathrm{UCC}|\Psi_\mathrm{UCC}}} = \frac{\braket{D_0|e^{-\hat \tau} \hat He^{\hat \tau}|D_0}}{\braket{D_0|e^{-\hat \tau}e^{\hat\tau}|D_0}}\\
               &=\braket{D_0|e^{-\hat \tau}\hat H e^{\hat \tau}|D_0}
\end{split}
\end{align}
The cluster coefficients are then usually obtained by setting
\begin{equation}
\frac{\partial \braket{E}_\mathrm{UCC}}{\partial t_\mathbf{i}} = 0
\end{equation}
for all $t_\mathbf{i}$ in the expansion.
\newline\newline
For $\bar H = e^{-\hat \tau}\hat H e^{\hat \tau}$, the CBH expansion 
\begin{align}
\label{eq:uccbch}
\begin{split}
\bar H = &\hat H + [\hat H, \hat T] + [\hat T^\dagger, \hat H]  \\
&+ \frac{1}{2} ([[\hat H, \hat T],\hat T] + [T^\dagger, [\hat T^\dagger, \hat H]] + [\hat H, [\hat T, \hat T^\dagger]]) + ...
\end{split}
\end{align}
no longer terminates at a finite order.
This is due to the presence of mixed terms such as $[\hat H, [\hat T, \hat T^
\dagger]]$ in \cref{eq:uccbch}, which have no termination point, \cite{Szalay1995} leading 
to infinite series for the energy and amplitude equations. Most implementations 
of UCC rely on some truncation of these expressions, either to a particular 
order in perturbation theory \cite{Bartlett1989} or to a particular power of $\hat \tau$.
\newline\newline
While this has not been the most common approach taken in the past, it is also possible\cite{Pal1984} to solve a set of projected UCC equations, 
\begin{equation}
\braket{D_\mathbf{i}|\hat H - E|\Psi_\mathbf{UCC}} = 0,
\end{equation}
or their similarity transformed counterpart,
\begin{equation}
\braket{D_\mathbf{i}|\bar H - E|D_0} = 0.
\end{equation}
Much like traditional coupled cluster and variational coupled cluster do not 
in general lead to the same energy or indeed wavefunction, we expect that 
projective and variational UCC will also generate different results. We will 
focus our attention on the projective method as it is naturally more amenable 
to the QMC algorithms described below. An example of the differences  between these methods can be found in section V.
\newline\newline
Deterministically solving the projected UCC equations also requires a
truncation of the exponential to make the computation tractable. In our case,
we have chosen to truncate at a given order of $\hat \tau$ and have found that
the results converge rapidly with increasing truncation order, as is shown
in one of the following sections. Using the truncated exponential, one can 
naively solve the residual equations iteratively, by {\color{black} starting 
from a Hartree--Fock wavefunction ($\forall \mathbf{i}:t_\mathbf{i} = 0$)} and 
taking steps of the form
\begin{equation}
t_\mathbf{i}(\beta+\delta\beta) = t_\mathbf{i}(\beta) - \delta \beta\braket{D_\mathbf{i}| \hat H -E_\mathrm{proj}|\Psi_\mathrm{UCC}},
\end{equation}
where $E$ is the current projected energy estimate, $E_\mathrm{proj} = \braket {D_0|\hat H|\Psi_\mathrm{UCC}}/\braket{D_0|\Psi_\mathrm{UCC}}$. We find that 
this approach converges well, albeit slowly. This could likely be improved by 
using more involved optimisation algorithms.
\newline\newline
Another approach to simplify the UCC \textit{ansatz} involves the Suzuki--Trotter 
decomposition\cite{Trotter1959, Suzuki1976} of the exponential,
\begin{equation}
e^{\hat \tau} = e^{\sum_\mathbf i \hat \tau_\mathbf i} \approx (\prod_\mathbf i e^{\frac{\hat \tau_\mathbf i}{\rho}})^\rho,
\label{eq:trot}
\end{equation}
where the index $\mathbf i$ in \cref{eq:trot} runs over all possible 
excitations. This approximation recovers the original \textit{ansatz} in the 
limit that $\rho \rightarrow \infty$, but recently it has been shown that a 
value of $\rho = 1$ is sufficient to give an exact parametrisation of the 
wavefunction, provided the operators are arranged in a particular order.\cite{Evangelista2019} 
Having replaced the full UCC \textit{ansatz} with a trotterized form, the 
resulting equations can then be solved either variationally or by 
projection, as discussed above.
\newline\newline

\section{Coupled Cluster Monte Carlo}

The CC equations can be equivalently written as
\begin{equation}
\braket{D_{\textbf i}|1-\delta\beta(
\hat H - E)|\Psi_{\mathrm{CC}}} = \braket{D_{\textbf i}|\Psi_{\mathrm{CC}}}
\end{equation}

and recast in an iterative form as\cite{Spencer2016}
\begin{equation}
t_{\textbf i}(\beta + \delta\beta) = t_{\textbf i}(\beta) - \delta\beta\braket
{D_{\textbf i}|\hat H-E|\Psi_{\mathrm{CC}}}
\label{eq:iter}
\end{equation}

The iterative equations describe the population dynamics of a set of particles 
in the Hilbert space. This dynamics is controlled by two processes, 
corresponding to the action of diagonal and off-diagonal Hamiltonian elements 
respectively: 
\begin{enumerate}
\item death/cloning of particles on $\ket{D_{\textbf i}}$.
\item spawning of a particle from 
$\ket{D_{\textbf i}}$ onto another  $\ket{D_{\textbf j}}$ coupled to it by 
the action of the Hamiltonian 
\end{enumerate}
A third process, annihilation, where pairs of particles of opposite 
signs on the same excitor are removed, is required to mitigate the 
sign problem in the algorithm.\cite{Booth2009} \textcolor{black}{These processes allow for the variation of the normalisation of the wavefunction, so one can write the CCMC wavefunction as $\Psi_\mathrm{CCMC} = N_0 e^{\sum_\mathbf{i} \frac{N_\mathbf{i}}{N_0} \hat a_\mathbf{i}}$, where $N_0$ corresponds to the population on the reference and $N_\mathbf{i}$ to the population on the excitor $\mathbf i$.}
\newline\newline
Such a calculation provides two estimators for the correlation energy of a 
system, which should agree once the population dynamics have reached a steady--state corresponding to the ground state wavefunction. Firstly, one can compute 
the instantaneous projected energy 
\begin{equation}E_{\mathrm{proj}} = \frac
{\braket{D_{0}|\hat H|\Psi_{\mathrm{CCMC}}}}{\braket{D_{0}|\Psi_{\mathrm{CCMC}}}}.
\end{equation}
Secondly, a shift $S$ is introduced to replace the unknown $E$ in 
\cref{eq:iter} and act as a population-control parameter. If it is allowed to 
vary such that a stable particle population is maintained,\cite{Booth2009} the  shift will converge onto the value of the correlation energy.
\newline\newline
The population dynamics described above can be performed stochastically, by
sampling the first two processed with probabilities given by the values of 
the relevant Hamiltonian terms. In the original implementation of CCMC, the 
sampling was carried out in the following way:\cite{Scott2017}

\begin{enumerate}
\item a cluster size $s$ is selected with probability 
\begin{equation}
p(s) = \frac{1}{2^{s+1}}
\end{equation}

\item a particular cluster of $s$ distinct excitors is selected with probability 
\begin{equation}
p(e|s) = s! \prod_{i=1}^s \frac{|N_i|}{|N_\mathrm{ex}|}
\label{eq:psel}
\end{equation}
where $N_\mathrm{ex}$ is the total population on excitors. The total selection 
probability is therefore 
\begin{equation}
p_\mathrm{sel}(e) = p(e|s)p(s)
\end{equation}
\item spawning from $\ket{D_\mathbf{i}}$ to $\ket{D_\mathbf j}$ creates a 
particle with probability 
\begin{equation}
p_\mathrm{spawn} = \delta \beta \frac{|w_e|}{n_\mathrm{a}p_\mathrm{sel}(e)}
\frac{|H_\mathbf{ij}|}{p_\mathrm{gen}},
\end{equation}
where $w_e$ is the total amplitude on the cluster which collapses to $\ket{D_
\mathbf{i}}$, $n_\mathrm{a}$ is the number of spawning attempts and 
$p_\mathrm{gen}$ is the probability of generating $\ket{D_\mathbf j}$
\item death occurs with probability 
\begin{equation}
p_\mathrm{death} = \delta \beta \frac{|w_e|}{n_\mathrm{a} p_\mathrm{sel}(e)}|H_
\mathbf{ii} - S|
\end{equation}
\end{enumerate}
More recently, an importance sampling based method has been developed,\cite{Scott2017} 
which allows the term $\frac{|w_e|}{n_\mathrm a p_\mathrm{sel}(e)}$
to be provably equal to 1 for all cluster sizes, thereby decreasing the time 
spent sampling large clusters that are unlikely to contribute.

Further improvements of the algorithm have been implemented, including an initiator approximation,\cite{Spencer2016} a linked approach,\cite{Franklin2016} efficient parallelisation\cite{Spencer2018} and excitation generators.\cite{Holmes2016, Neufeld2019}

\subsection{Variants of stochastic coupled cluster}
\textcolor{black}{What we have described above is the initial, unlinked implementation of Coupled Cluster Monte Carlo. Since then a linked CCMC formalism\cite{Franklin2016}, as well as a diagrammatic version of CCMC\cite{Scott2019} have been developed.}

\textcolor{black}{In linked CCMC, one follows a similar procedure to that described above, but sampling the action of the similarity transformed Hamiltonian $\bar H = e^{-\hat T} \hat H e^{\hat T}$, rather than $\hat H$. This involves sampling the four commutators in \Cref{eq:cccbh}, which requires some significant changes to the selection algorithm described above.\cite{Franklin2016} Firstly, and helpfully, clusters may have at most four excitors. Secondly, excitors which act on some of the same orbitals may give a non-zero contribution to the energy and must therefore be considered. Thirdly, once a particular set of excitors has been selected, all possible orderings of the excitors and the Hamiltonian in the commutator must be considered simultaneously, to maintain the benefit of only sampling connected terms. Finally, for spawning and death, one must build the relevant terms of the similarity transformed Hamiltonian. The final three steps all induce significant added complexity to the CCMC algorithm, however this is offset by the reduction in size of the expansion that needs to be considered. 
}

\textcolor{black}{In contrast, in diagrammatic CCMC, one considers the similarity transformed normal ordered Hamiltonian,
\begin{align}
\begin{split}
\bar H_\mathrm{N} &= e^{-\hat T} H_\mathrm{N} e^{\hat T} \\
&= \hat H_\mathrm{N} + {\contraction{}{H_\mathrm{N}}{}{T} H_\mathrm{N}T} +{\contraction{}{H_\mathrm{N}}{}{T} \contraction{}{H_\mathrm{N}}{T}{T} H_\mathrm{N}TT} +  {\contraction{}{H_\mathrm{N}}{}{T} \contraction{}{H_\mathrm{N}}{T}{T} \contraction{}{H_\mathrm{N}}{TT}{T} H_\mathrm{N}TTT} + {\contraction{}{H_\mathrm{N}}{}{T} \contraction{}{H_\mathrm{N}}{T}{T} \contraction{}{H_\mathrm{N}}{TT}{T}\contraction{}{H_\mathrm{N}}{TTT}{T}H_\mathrm{N}TTTT}
\end{split}
\label{eq:cbh-diag}
\end{align}
for which the CBH expansion reduces to the connected contributions shown above, which can be represented in diagrammatic form.\cite{Shavitt2009} Since the cluster operator $\hat T$ only appears to the right of the Hamiltonian in \Cref{eq:cbh-diag}, these terms are simpler to sample than the corresponding commutators in \Cref{eq:cccbh}. The cluster amplitudes can be found by sampling the update equation
\begin{equation}
t_{\textbf i}(\beta + \delta\beta) = t_{\textbf i}(\beta) - \delta\beta\braket
{D_{\textbf i}|\bar H_\mathrm{N}|D_0}.
\end{equation}
This can be done by selecting particular diagrams relevant to the second term on the right hand side and computing their contributions, as described in Ref. \onlinecite{Scott2019}. While the selection scheme required is still more involved than the one employed in unlinked CCMC, it avoids some of the complications of the linked formalism, while also more strongly imposing connectivity constraints on the considered diagrams, leading to promising performance improvements over the unlinked formalism.}

\textcolor{black}{The main benefit these approaches have over the unlinked CCMC formalism comes from their ability to encode the terminating CBH expansion of the similarity transformed Hamiltonian, therefore guaranteeing that fewer clusters need to be considered. For UCC this expansion is non-terminating and therefore the benefits of employing the linked or diagrammatic formalisms are much diminished, while the additional computational complexities remain. Therefore, in our initial stochastic approach to the UCC problem, we will focus on the original, unlinked formalism, whose relative computational simplicity provides an ideal testing ground for the feasibility of such an endeavour.}

\section{Stochastic Unitary Coupled Cluster}

The stochastic implementation of UCC (herein referred to as UCCMC) is based 
around the same principles as that of traditional CCMC.  One must select a 
cluster amplitude and allow it to undergo spawning, death and annihilation. 
Modifications that must be made to the original algorithm to accommodate for 
the change in the cluster operator are detailed below, for both full and 
trotterized UCCMC.
\subsection{Full unitary coupled cluster}
The presence of deexcitation operators $\hat T^\dagger$ in the full UCC
\textit{ansatz} substantially changes the structure of the allowed clusters. 
In traditional CCMC, the largest allowed excitation level of any considered 
cluster is $n+2$, where $n$ is the considered truncation level, as this is the 
highest order excitation that couples to the CCMC wavefunction through the 
Hamiltonian. As such a cluster could be formed from at most $n+2$ single excitors, this 
is also the largest size of cluster one must consider. In UCC however, the 
inclusion of deexcitation operators can lower the overall excitation level of 
the cluster while increasing its size, so this heuristic no longer holds.
Therefore, in principle, for an implementation of UCCMC, one must consider 
clusters of up to infinite size. However, as in the deterministic case, one can 
truncate the expansion to a finite size of cluster.  Indeed, in the original implementation
of CCMC in HANDE-QMC,\cite{HANDE2018} the maximum allowed cluster size 
is 12, due to technical limitations in computing a factorial. This size 
limitation has been preserved, and we have found that even for larger 
systems like N$_2$, valid clusters of size 12 are sampled extremely rarely, so 
increasing the polynomial truncation level would not improve the precision of our current 
algorithm. Having taken this into consideration, the selection scheme for UCCMC 
is as follows:
\begin{enumerate}
\item select a cluster size $s$ with probability \mbox{$p(s) = \frac{1}{2^{s+1}}$}. Other distributions, such as the uniform distribution or distributions where $p(s)$ increases with $s$, have been attempted here, with little to no effect on the quality of the results. However, as in the case of CCMC, it is possible that tweaking the selection probabilities may improve the efficiency of the algorithm.
\item for all but the first excitor in the cluster, decide with probability 
$\frac{1}{2}$ whether it will be an excitation or deexcitation operator
\item a particular cluster is selected as before, with probability given by 
\cref{eq:psel}
\end{enumerate}

Having selected the cluster, it undergoes spawning and death as before. The 
final aspect one must be careful of is the normalisation of the HF reference. 
In the case of CCMC and FCIQMC, the population on the reference was equal to 
the wavefunction projection onto the reference determinant, $\braket{D_0|\Psi_
\mathrm{CCMC}} = N_0$. Consequently, other excitor populations can be 
normalised relative to this to give an intermediately normalised wavefunction 
as naturally arises from the exponential form of the \textit{ansatz} and the 
correlation energy can be computed as
\begin{equation}
E_\mathrm{proj} = \sum_{i\neq 0} \frac{N_\mathbf i^\mathrm{CI}}{N_0}H_{\mathbf i 0}
\label{eq:proj}
\end{equation}
where $N_\mathbf{i}^\mathrm{CI}$ is the equivalent CI population on a 
determinant, obtained by sampling all combinations of different-sized
clusters collapsing onto that determinant.
However, for UCCMC, it is trivial to show that $\braket{D_0|\Psi_\mathrm{UCC}} 
\neq 1$. Consider the Taylor expansion of the exponential
\begin{equation}
e^{\hat T - \hat T^\dagger} = 1 + \hat{T} - \hat T^\dagger + \frac{1}{2}(\hat T \hat T - \hat T \hat T^\dagger - 
\hat T^\dagger \hat T + \hat T^\dagger \hat T^\dagger) + ...
\end{equation}
If we further expand $\hat T=\sum_\mathbf{i} t_\mathbf{i} \hat a_\mathbf{i}$, where 
$\mathbf i$ indexes over all allowed excitors and $\hat a_\mathbf i$ is 
the corresponding excitation operator, then the mixed term $\hat T^\dagger \hat 
T$ becomes
\begin{equation}
\hat T^\dagger \hat T = \sum_\mathbf{i} t_\mathbf{i}^2 \hat a_\mathbf{i}^\dagger \hat a_\mathbf{i} + \sum_{\mathbf{i} \neq \mathbf{j}}t_\mathbf{i} t_\mathbf{j} \hat a_\mathbf{i}^\dagger \hat a_\mathbf{j}
= \sum_\mathbf{i}t_\mathbf{i}^2 \hat I + \sum_{\mathbf{i} \neq \mathbf{j}}t_\mathbf{i} t_\mathbf{j} \hat a_\mathbf{i}^\dagger \hat a_\mathbf{j}
\end{equation}
Therefore, this mixed squared term contributes to the projection onto the 
reference determinant with $\frac{1}{2}\sum_\mathbf{i}t_\mathbf{i}^2$. 
Similarly, any even powered term of the form
$\hat T^\dagger \hat T \hat T^\dagger \hat T ... \hat T^\dagger \hat T$ will 
have a similar contribution, leading to
\begin{equation}
\braket {D_0 |\Psi_\mathrm{UCC}} = 1 + \sum_\mathbf{i}\Big(\sum_{p=1}^\infty (-1)^p \frac{t_\mathbf{i}^{2p}}{2p!}\Big)
\end{equation}
\textcolor{black}{Considering a UCCMC wavefunction $\Psi_\mathrm{UCCMC} = N_0 e^{\sum_\mathbf{i} \frac{N_\mathbf{i}}{N_0}(\hat a_\mathbf{i} - \hat a_\mathbf{i}^\dagger)}$,}
\begin{equation}
\braket {D_0 |\Psi_\mathrm{UCCMC}} = N_0\Big\{1 + \sum_\mathbf{i}\Big[\sum_{p=1}^\infty (-1)^p \Big(\frac{N_\mathbf{i}}{N_0}\Big)^{2p}\Big]\Big\}
\end{equation}
Therefore, when one normalises the cluster coefficients relative to $N_0$, the 
projected energy becomes:
\begin{equation}
E_\mathrm{proj} = \sum_{i\neq 0} \frac{N_\mathbf i^\mathrm{CI}  H_{\mathbf i 0}}{N_0\Big\{1 + \sum_\mathbf{i}\Big[\sum_{p=1}^\infty (-1)^p \Big(\frac{N_\mathbf{i}}{N_0}\Big)^{2p}\Big]\Big\}}
\label{eq:proj_u}
\end{equation} 
The denominator does not need to be explicitly computed and can be sampled 
stochastically during the course of the calculation, concurrently with $N_0$ 
and $N_\mathbf i^\mathrm{CI} H_{\mathbf i 0}$.

\subsection{Trotterized unitary coupled cluster}
Consider the trotterized UCC \textit{ansatz} with $\rho = 1$. 
\begin{equation}
\ket{\Psi_\mathrm{tUCC}} = \prod_\mathbf i e^{\hat \tau_\mathbf i}\ket{D_0}
\end{equation}
A particular term $e^{\hat \tau_\mathbf i}$ can be expanded as
\begin{align}
\begin{split}
e^{\hat \tau_\mathbf i} &= 1 + t_\mathbf{i}(\hat a_\mathbf i - \hat a_\mathbf{i}^\dagger) + \frac{t_\mathbf{i}^2}{2!} (\hat a_\mathbf{i} \hat a_\mathbf{i} - \hat a_\mathbf{i}\hat a_\mathbf{i}^\dagger 
- \hat a_\mathbf{i}^\dagger \hat a_\mathbf{i} + \hat a_\mathbf{i}^\dagger \hat a_\mathbf{i}^\dagger) \\
&+ \frac{t_\mathbf{i}^3}{3!} (\hat a_\mathbf{i} \hat a_\mathbf{i}\hat a_\mathbf{i} - \hat a_\mathbf{i}\hat a_\mathbf{i}\hat a_\mathbf{i}^\dagger 
- \hat a_\mathbf{i}\hat a_\mathbf{i}^\dagger \hat a_\mathbf{i} - \hat a_\mathbf{i}^\dagger \hat a_\mathbf{i}\hat a_\mathbf{i} + \hat a_\mathbf{i}\hat a_\mathbf{i}^\dagger \hat a_\mathbf{i}^\dagger \\
&+ \hat a_\mathbf{i}^\dagger \hat a_\mathbf{i}\hat a_\mathbf{i}^\dagger 
+ \hat a_\mathbf{i}^\dagger\hat a_\mathbf{i}^\dagger\hat a_\mathbf{i} - \hat a_\mathbf{i}^\dagger\hat a_\mathbf{i}^\dagger\hat a_\mathbf{i}^\dagger ) + ...
\end{split}
\end{align}
Any term that that applies two of the same operator sequentially vanishes when 
applied to any wavefunction, so this can be rewritten
\begin{align}
\begin{split}
e^{\hat \tau_\mathbf i} &= 1 + t_\mathbf{i}(\hat a_\mathbf i - \hat a_\mathbf{i}^\dagger) + \frac{t_\mathbf{i}^2}{2!} (- \hat a_\mathbf{i}\hat a_\mathbf{i}^\dagger 
- \hat a_\mathbf{i}^\dagger \hat a_\mathbf{i})  \\
&+ \frac{t_\mathbf{i}^3}{3!} (- \hat a_\mathbf{i}\hat a_\mathbf{i}^\dagger \hat a_\mathbf{i} + \hat a_\mathbf{i}^\dagger \hat a_\mathbf{i}\hat a_\mathbf{i}^\dagger) + ...
\end{split}
\end{align}
Consider applying $e^{\hat \tau_\mathbf i}$ to an arbitrary single determinant 
wavefunction $\ket{\Psi}$. There are then three possibilities:
\newline\newline
\textbf{CASE 1:} $\hat a_\mathbf{i}^\dagger\ket{\Psi} = 0$ and $\hat a_\mathbf{i}\ket{\Psi} \neq 0$ 
\begin{align}
\begin{split}
e^{\hat \tau_\mathbf i}\ket{\Psi} &= \sum_{p=0}^\infty (-1)^p\Big( \frac{t_\mathbf{i}^{2p}}{2p!}\hat I + \frac{t_\mathbf i^{2p+1}}{(2p+1)!}\hat a_\mathbf{i}\Big)\ket{\Psi} \\
&= \cos(t_\mathbf i)\ket{\Psi} + \sin(t_\mathbf i) \ket{\Psi_\mathbf{i}} 
\end{split}
\end{align}
where $\ket{\Psi_\mathbf{i}}$ is the result of applying the excitation to $\ket{\Psi}$.
\newline\newline
\textbf{CASE 2:} $\hat a_\mathbf{i}^\dagger\ket{\Psi} \neq 0$ and $\hat a_\mathbf{i}\ket{\Psi} = 0$ 
\begin{align}
\begin{split}
e^{\hat \tau_\mathbf i}\ket{\Psi} &= \sum_{p=0}^\infty (-1)^p\Big( \frac{t_\mathbf{i}^{2p}}{2p!}\hat I - \frac{t_\mathbf i^{2p+1}}{(2p+1)!}\hat a_\mathbf{i}^\dagger\Big)\ket{\Psi} \\
&= \cos(t_\mathbf i)\ket{\Psi} - \sin(t_\mathbf i) \ket{\Psi^\mathbf{i}} 
\end{split}
\end{align}
where $\ket{\Psi^\mathbf{i}}$ is the result of applying the deexcitation to $
\ket{\Psi}$.
\newline\newline
\textbf{CASE 3:}  $\hat a_\mathbf{i}^\dagger\ket{\Psi} = 0$ and $\hat a_\mathbf{i}\ket{\Psi} = 0$
\begin{align}
\begin{split}
e^{\hat \tau_\mathbf i}\ket{\Psi} = \ket{\Psi}  
\end{split}
\end{align}

To translate this into a stochastic algorithm, an ordering of excitors is 
defined. The consequences of this choice are discussed in the following 
section. By default, a choice consistent with Ref. \onlinecite{Evangelista2019} is used, 
with excitors applied in decreasing order of highest orbital excited from and 
increasing order of excitation level. For each excitor present in the 
wavefunction, the algorithm assesses which of the cases listed above is 
appropriate. If the excitor cannot be applied (case 3), the next excitor is 
considered. If the excitor can be applied, the probability of doing so is 
computed as 
\begin{equation}
p_\mathrm{excit} = \frac{|\sin(t)|}{|\sin(t)| + |\cos(t)|} 
\end{equation}
With probability $p_\mathrm{excit}$, the excitor is applied and the cluster 
amplitude is multiplied by $\sin(t)$. With probability $1-p_\mathrm{excit}$, 
the operator is not applied and the cluster amplitude is multiplied by $\cos(t)
$. Once a cluster is selected in this way it undergoes the same spawning, death 
and annihilation steps as in the traditional CCMC algorithm. As in the case of 
full UCCMC, this \textit{ansatz} modifies the normalisation of the wavefunction
with respect to the reference. In this case,
\begin{equation}
\braket{D_0|\Psi_\mathrm{tUCC}} = \prod_\mathbf{i} \cos(t_\mathbf{i}).
\end{equation}
Therefore, in the stochastic representation,
\begin{equation}
t_\mathbf{i} = \frac{N_\mathbf i}{N_0}
\end{equation}
and
\begin{equation}
E_\mathrm{proj} = \sum_{i\neq 0} \frac{N_\mathbf i^\mathrm{CI} H_{\mathbf i 0}}{N_0\prod_{\mathbf{i}}\cos\Big(\frac{N_\mathbf i}{N_0}\Big)}
\label{eq:proj_t}
\end{equation}

\subsection{Energy estimators}
\label{sec:energ}
\textcolor{black}{
To end this section, we go through the available energy estimators and how they are obtained in each case.
\begin{enumerate}
\item The shift $S$ is output directly from the calculation at each report loop. The final value is obtained by a reblocking analysis.\cite{Flyvbjerg1989}
\item The projected energy $E_\mathrm{proj}$, given by \cref{eq:proj}, \cref{eq:proj_u} or \cref{eq:proj_t}, is computed as a ratio of the average values of the numerator and denominator. Estimates of these are output at each report loop and averages obtained by a reblocking analysis.
\item The expectation value of the energy $\braket{E}$ is not generally computed for QMC methods, or indeed for conventional CC. However, to obtain a fairer comparison with the variational UCC method, we have implemented the following procedure. At the end of each calculation a list of average cluster coefficients is output. These are used to generate the corresponding CI wavefunction and the expectation value of the energy is computed. This value is averaged over 5 independent calculations to obtain the final estimate.
\end{enumerate}
}
\section{Numerical Results}
\label{sec:num}
\textcolor{black}{The following section presents results obtained using a development version of HANDE-QMC.\cite{HANDE2018} Values for the projected energy are obtained from a reblocking analysis,\cite{Flyvbjerg1989} while the expectation value of the energy is obtained as described in \Cref{sec:energ}. In all cases, errorbars correspond to a single standard deviation, $\sigma$. In the case of FCIQMC and CCMC, the error bars decrease with imaginary time as $\sigma \propto \sqrt{\beta^{-1}}$ and the same behaviour is observed in UCCMC (see Supplementary Information).}
\subsection{Two-electron systems}
\begin{figure}[h!]
\subfloat{\includegraphics[width=0.5\textwidth, center]{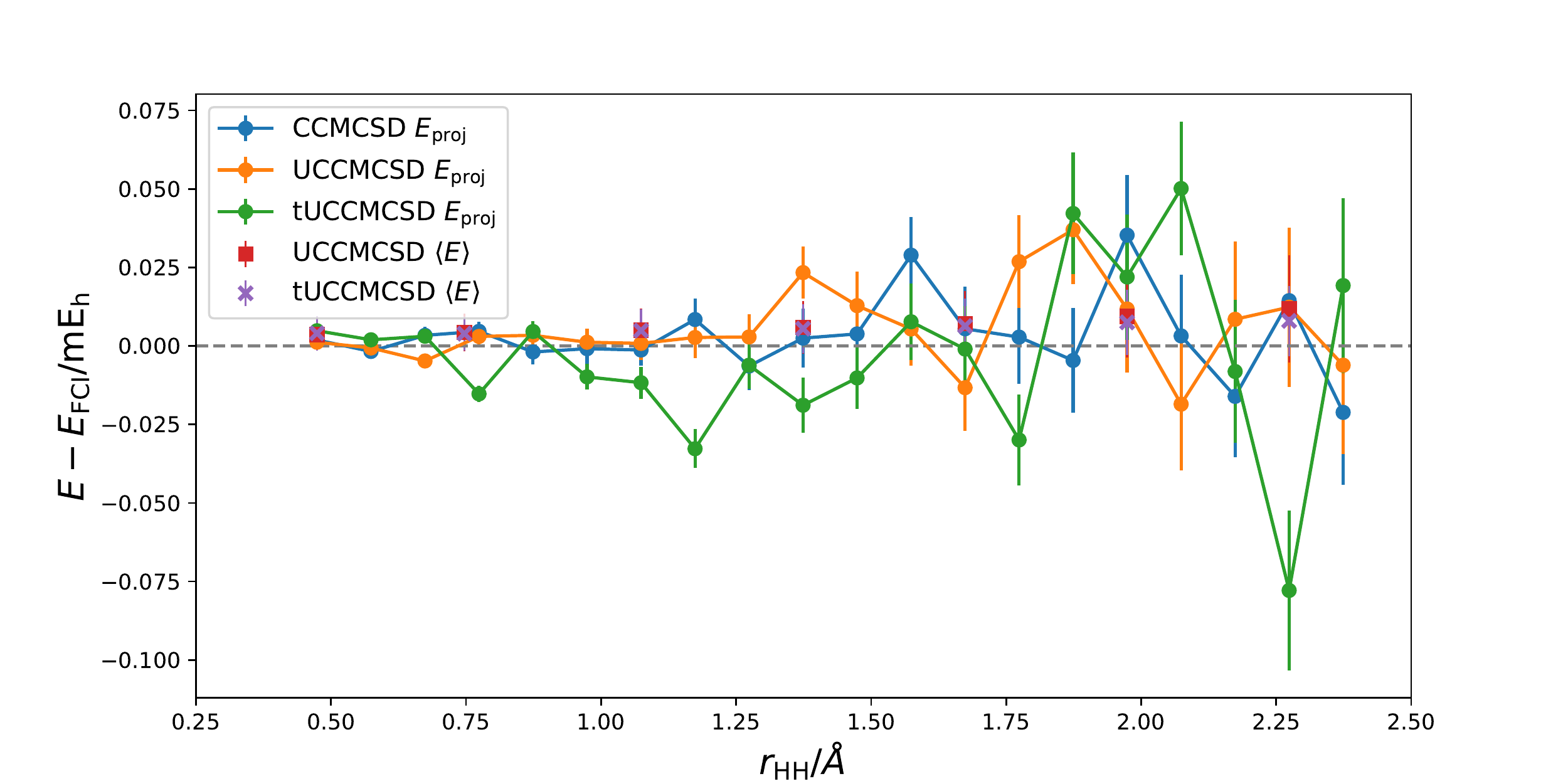}}
\caption{\raggedright \small Error in stochastic coupled cluster and unitary coupled cluster energy 
estimates relative to the exact UCCSD (FCI) energy 
for H$_2$ in a STO-3G basis set.}
\label{fig:h2-sto-e}
\end{figure}

\begin{figure}[h]
\subfloat{\includegraphics[width=0.5\textwidth, center]{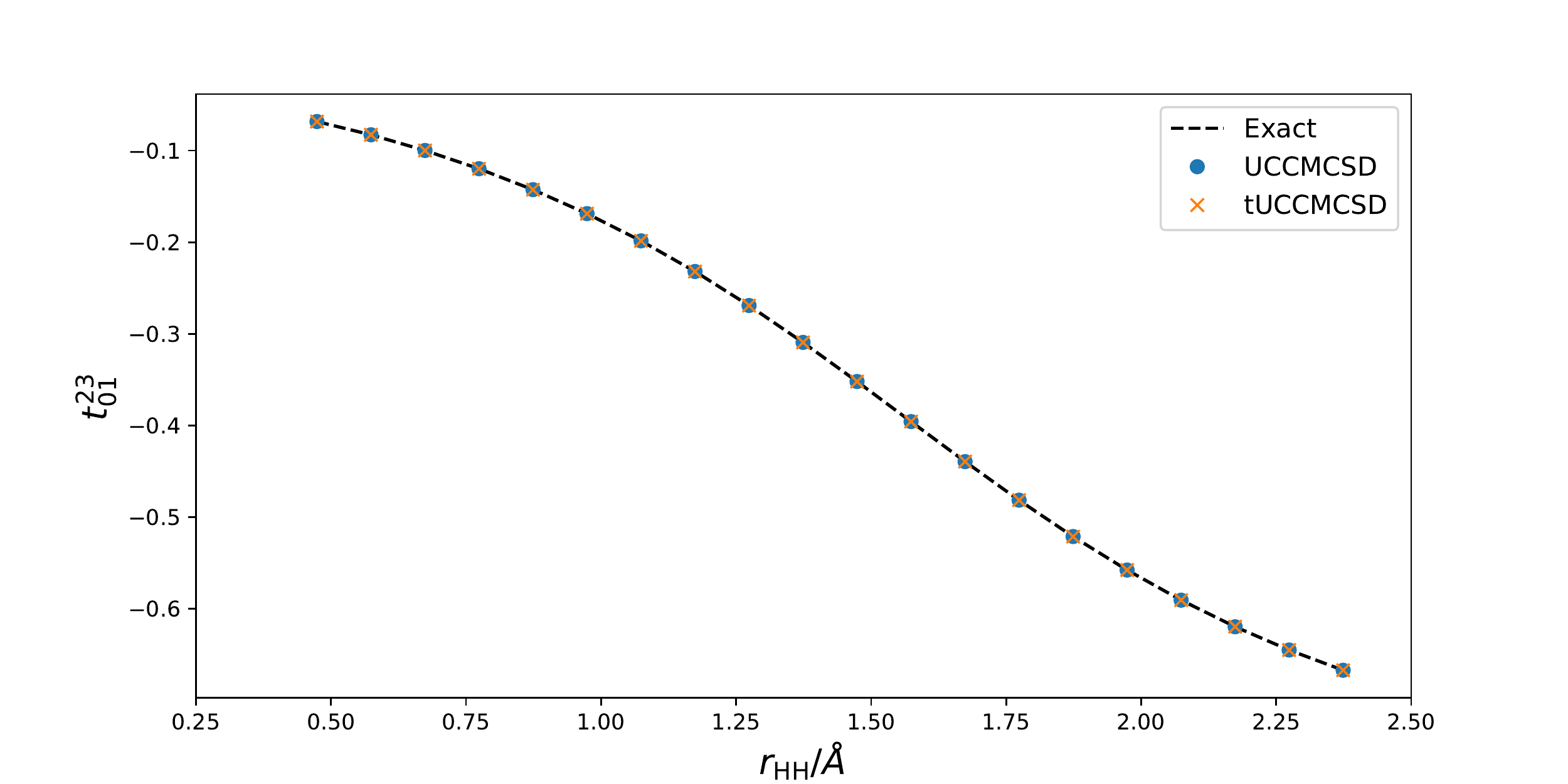}}
\caption{\raggedright \small Error in stochastic UCCSD and tUCCSD estimates of the coefficient $t$ on $
\ket{D_{01}^{23}}$ for H$_2$ in a STO-3G basis set.}
\label{fig:h2-sto-c}
\end{figure}

The natural starting point for the investigation of \mbox{UCCMC} energies and 
wavefunctions is H$_2$. As it only has two electrons, UCCSD should be 
exact in this case. Particularly, in the STO-3G basis set,\cite{Hehre1969} only 
two determinants contribute to the wavefunction, which may be written 
in the UCC form as
\begin{equation}
\ket{\Psi_\mathrm{UCC}} = e^{t \hat a_{12}^{34}}\ket{D_0}
= \cos(t)\ket{D_0} + \sin(t)\ket{D_{01}^{23}}
\end{equation}
It is therefore trivial to analytically solve either the variational or the projected equations 
to obtain $t$ and the exact UCC energy, which is in this case 
equivalent to FCI. Stochastic estimates of the energy and the coefficient $t$ have been 
obtained using both full UCCMCSD and trotterized UCCMCSD (tUCCMCSD). The 
differences relative to the exact values are given in 
\Cref{fig:h2-sto-e} and \Cref{fig:h2-sto-c}. Both show good agreement 
between the stochastic estimates and the exact values. In particular, 
the expectation-value estimator for the energy, which has been implemented 
for a fairer comparison to conventional UCC, shows excellent agreement 
with the exact energy.
\newline

\begin{figure}[h]
\subfloat{\includegraphics[width=0.45\textwidth, center]{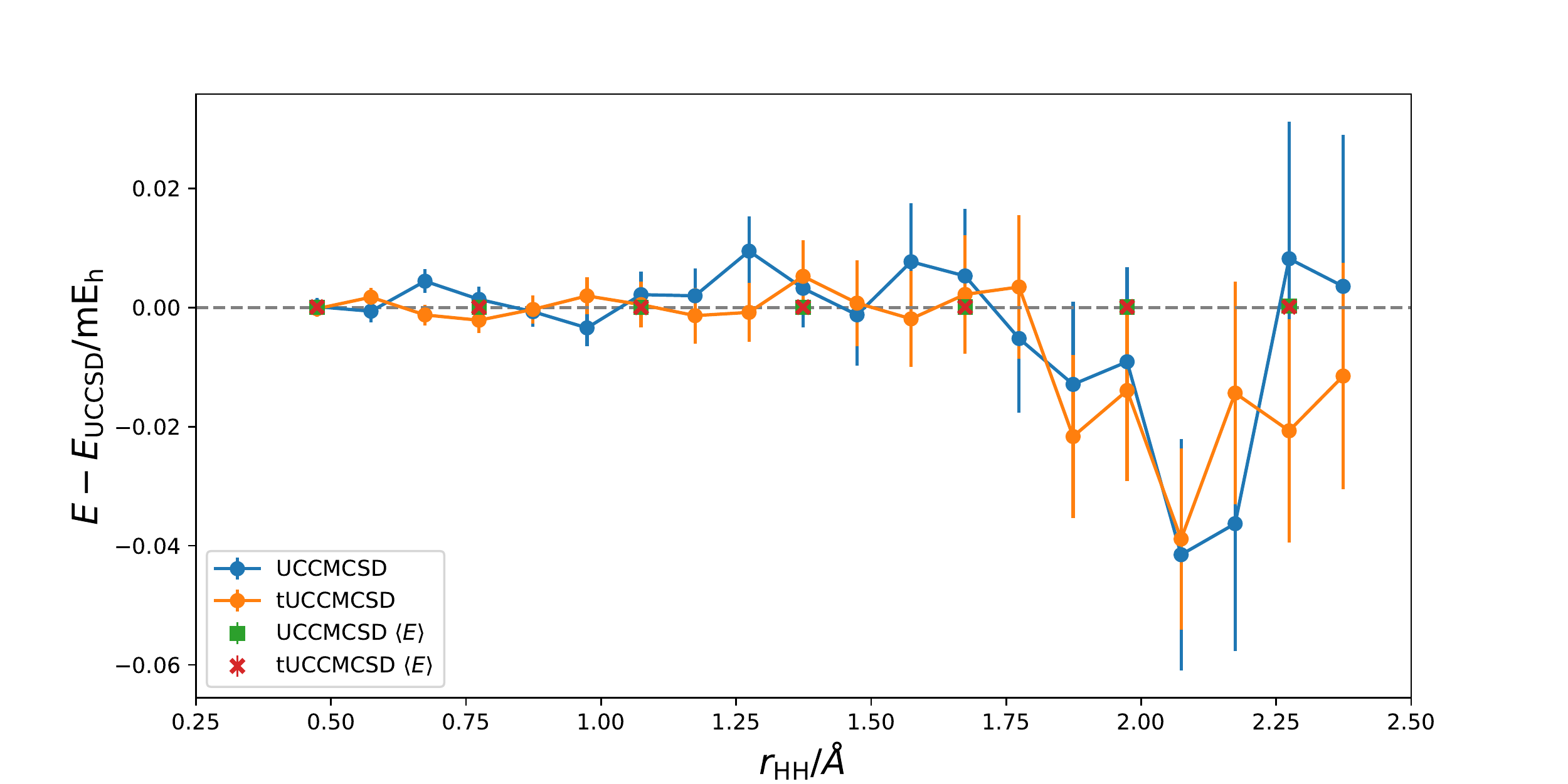}}
\caption{\raggedright \small Error in stochastic UCCSD and tUCCSD energy estimates relative 
to the exact UCCSD (FCI) energy for H$_2$ in a 6-31G basis set.}
\label{fig:h2-631-e}
\end{figure}

Increasing the basis to 6-31G\cite{Ditchfield1971} leads to a system with 7 symmetry-allowed 
excited determinants. \Cref{fig:h2-631-e} shows the energies of stochastic UCC 
approaches relative to the exact value. The agreement is good, but one can 
observe the increase in the size of the error bars of the projected energy as the bond length increases. 
This is caused by the increase in static correlation as the molecule approaches 
dissociation. In simple cases such as this, the size of the error bars can be 
decreased by simply running longer calculations at higher bond lengths. 

\begin{figure}[h!]
\subfloat{\includegraphics[width=0.5\textwidth, center, trim=1.5cm 5cm 3cm 7.5cm,clip=true]{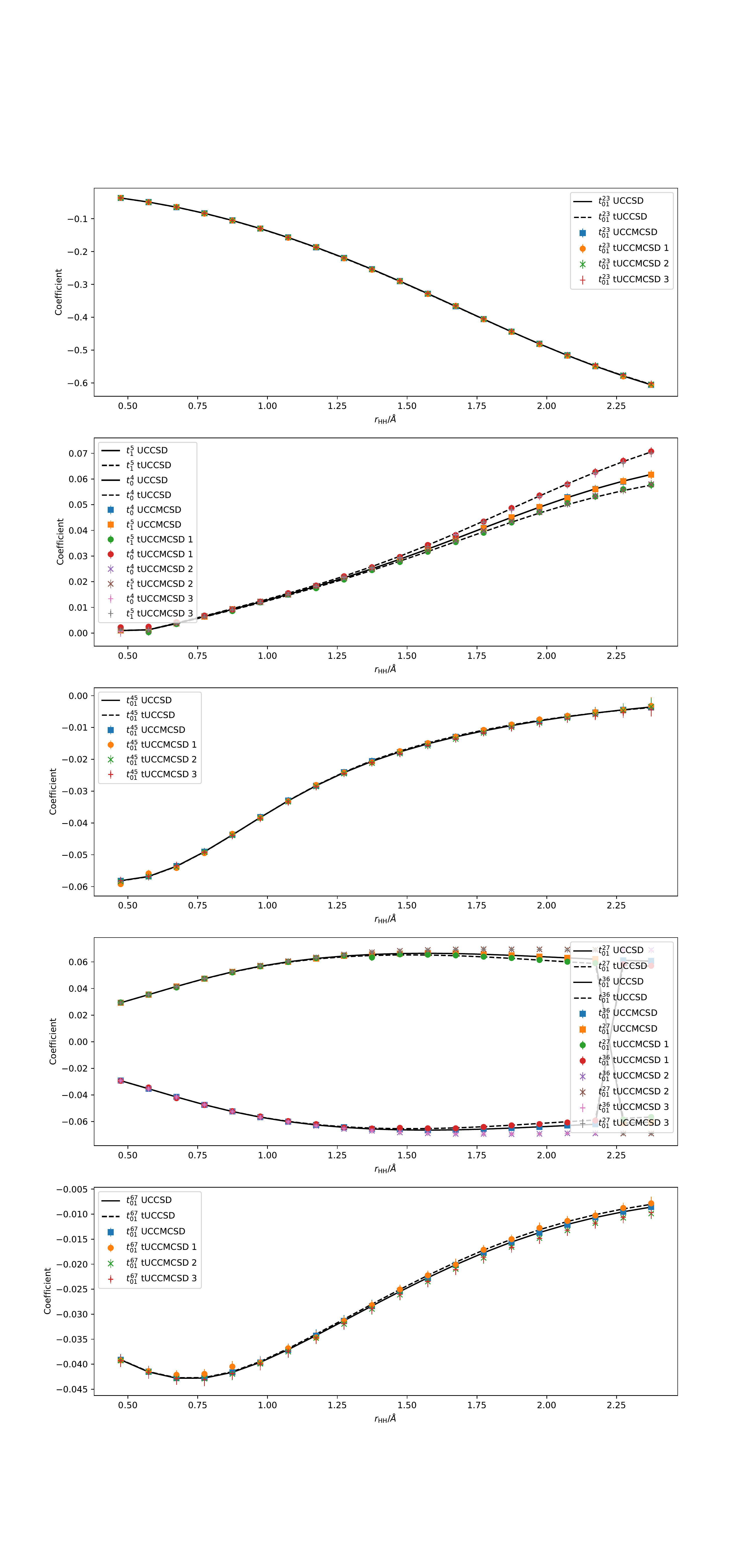}}
\caption{\raggedright \small UCCMCSD and tUCCMCSD coefficients for H$_2$ in a 6-31G 
basis set.
\color{black} The first set of tUCCMCSD results use an order consistent with\cite{Evangelista2019}, giving
$\ket{\Psi_\mathrm{tUCCSD}} = e^{\hat T_0^4}e^{\hat T_{01}^{23}} e^{\hat T_{01}^{43}} e^{\hat T_{01}^{25}} 
e^{\hat T_{01}^{45}} e^{\hat T_{01}^{67}} e^{\hat T_{1}^{5}}\ket{D_0}$. The second and third set use the alternative orders with single excitations applied first and last respectively. The orbital labels are chosen to reflect the physical nature of the orbitals, rather than the energy. Consequently, at $0.895 \leq r_\mathrm{HH}\leq 1.395$, spin-orbitals 4 and 5 are higher in energy than 6 and 7.}
\label{fig:h2-631-c}
\end{figure}

The coefficients obtained from UCCMCSD are also in good agreement with the 
ones obtained by solving the projected UCCMCSD equations deterministically, 
as can be seen in \Cref{fig:h2-631-c}. It is worth noting that in both cases the 
results were obtained with an expansion truncated at $\hat \tau^{12}$, but 
deterministic tests suggest that convergence with respect to polynomial order is reached at $\hat\tau^4$. 
For tUCCMCSD however, the ordering dependence becomes obvious. In UCC, the 
coefficients on spin-flipped excitors are equal up to a sign change. However, 
when using the order of Evangelista \textit{et al}\cite{Evangelista2019} (see 
\Cref{fig:h2-631-c} caption), these excitors acquire different amplitudes. An 
alternative ordering, where all single excitations are applied after all double 
excitations, recovers the equivalence of spin-flipped pairs, however these 
coefficients do not necessarily correspond to the UCC values. Clearly, the 
ordering of excitors in the Trotter expansion is a significant parameter of 
such a calculation and must be chosen carefully.  Further, from 
\Cref{fig:h2-631-c}, the observation can be made that the default ordering used 
for tUCCMC ensures one coefficient of each pair agrees with the full UCCMC 
result, while the other is modified by the ordering. This is observed in larger 
systems as well (see Supplementary Information).

\FloatBarrier
\subsection{Beyond two electrons}
Taking a small step away from the simplicity of two-electron systems, we look at LiH in a STO-3G basis set, which has 4 electrons in 12 spin-orbitals. 

\begin{figure}[h]
\subfloat{\includegraphics[width=0.5\textwidth, center]{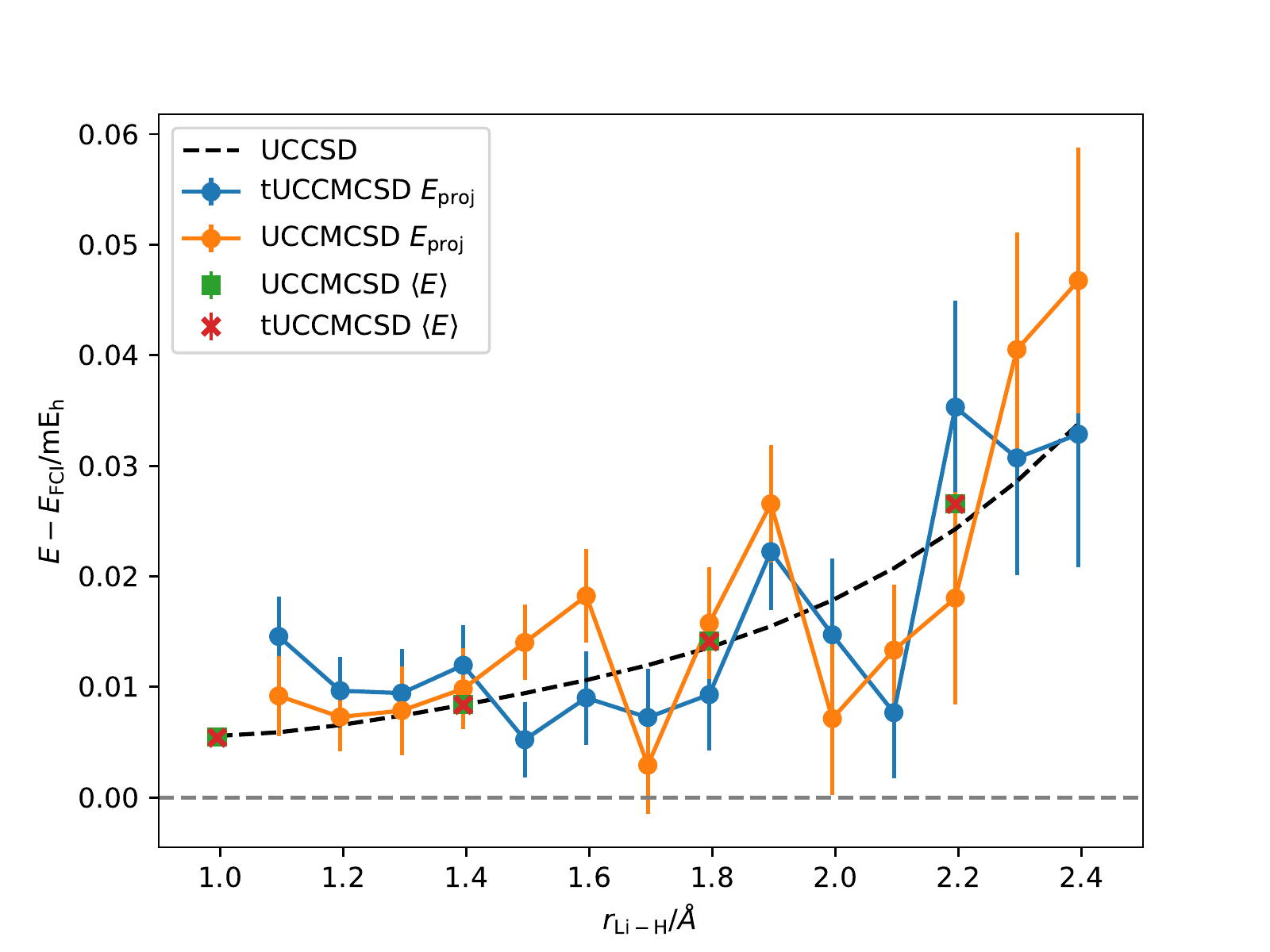}}
\caption{\raggedright\small Error in stochastic UCCSD energy 
estimates relative to the exact FCI energy for LiH in a STO-3G basis set.}
\label{fig:lih-sto-a-e}
\end{figure}

For LiH with all electrons correlated, UCCSD is no longer expected to be exact, and indeed the deterministic projected results show an increasing deviation from FCI as the bond length increases (see \Cref{fig:lih-sto-a-e}). 
The energies from stochastic UCCSD agree well with those from 
its deterministic counterpart, as can be seen from \Cref{fig:lih-sto-a-e}. 
Coefficients show similar behaviour to that observed for two-electron systems, 
with spin-flipped coefficients acquiring different values in tUCCMCSD (see Supplementary Information).

While small systems like H$_2$ and LiH are good models to test the fundamental
behaviour of new algorithms, they are hardly representative of the kind
of problems of interest in electronic structure today. To approach these, 
we will look at the dissociation of the nitrogen molecule, 
in which a triple bond must be broken. The system is therefore 
characterised by important strong correlation effects, which cause failures of 
both conventional and stochastic CC methods at large bond-lengths.
\cite{Chan2004}

\begin{figure}[h]
\subfloat{\includegraphics[width=0.5\textwidth, center]{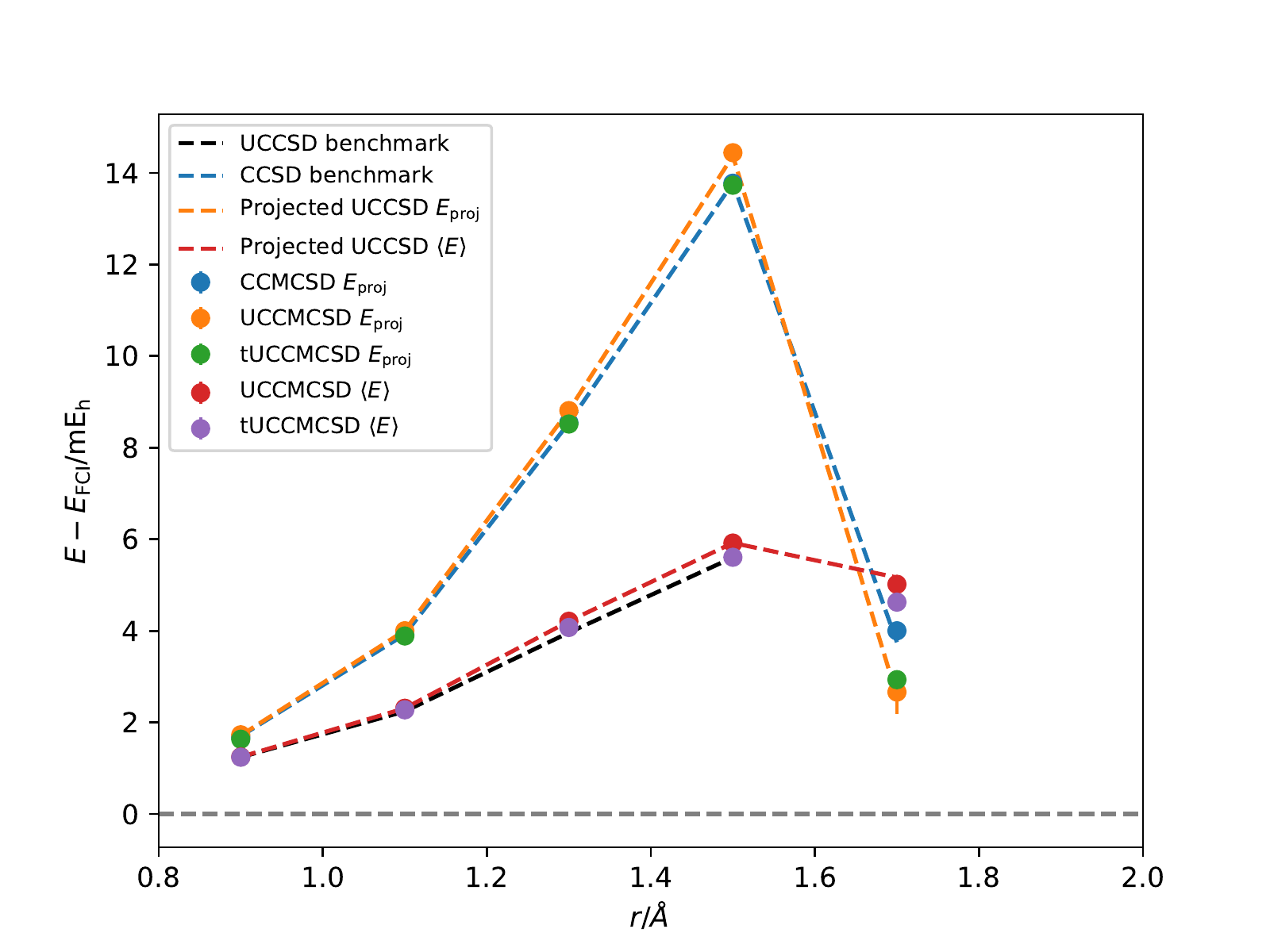}}
\caption{\raggedright \small Error in stochastic CCSD and UCCSD energy estimates relative to the FCI energy for N$_2$ in a STO-3G basis set. While the $E_\mathrm{proj}$ estimator for UCCMCSD and tUCCMCSD approaches that for CCSD, the expectation value $\braket{E}$ remains close to the variational UCCSD value. CCSD and UCCSD benchmark values are from Ref. \onlinecite{Cooper2010}. }
\label{fig:n2-sto-e}
\end{figure}

Variational UCCSD has been benchmarked for this system by Cooper and Knowles,\cite{Cooper2010} showing an improvement over traditional CCSD. By comparison, 
using the projected energy estimator from projected UCCSD (or tUCCSD) gives 
results that are very similar to CCSD. However, if one computes the expectation 
value of the energy instead, this recovers almost all of the correlation energy 
of the variational approach, as can be seen from \Cref{fig:n2-sto-e}. 
Additionally, the method converges beyond the point reported in Ref. 
\onlinecite{Cooper2010}. At bond lengths beyond those shown in \Cref{fig:n2-sto-e}, 
the stochastic method is difficult to converge and the projected energy dips 
below the FCI results, as it does for traditional CCSD, however the expectation 
value of the energy remains variationally above the FCI result.

\textcolor{black}{The N$_2$ system is sufficiently large to study the efficiency of UCCMC relative to traditional CCMC. We find that both the population plateau and the convergence of the projected energy estimator with imaginary time behave very similarly between CCMCSD, UCCMCSD and tUCCMCSD (see Supplementary Information). }

\textcolor{black}{We also investigate the effect the polynomial truncation of the UCCSD expansion has on the quality of the obtained energy estimators. Consider truncating the exponential expansion at a polynomial order $o$:
\begin{equation}
e^{\hat\tau} \approx \sum_{i=0}^o \frac{1}{i!}\hat\tau^i
\end{equation}
\Cref{tab:conv} gives the values of the energy computed using the projected UCCSD method truncated at orders $o=2-12$. One finds that system, $o=4$ provides a sub miliHartree approximation of the final result and $o=8$ appears converged to within $10^{-8}\mathrm{E_h}$. Therefore we are confident that the truncation at $o=12$ generally used in our calculations does not introduce any significant error into the results, in either the deterministic or stochastic case.}

\begin{table}
\begin{small}
\begin{tabular}{|c|c|c|}
\hline
\textbf{Truncation order $o$}&\textbf{$E_\mathrm{proj}/\mathrm{E_h}$}&\textbf{$\braket{E}/\mathrm{E_h}$}\\
\hline
2&-0.21\textcolor{red}{526093} &-0.22\textcolor{red}{010447}\\
3&-0.21\textcolor{red}{712594}&-0.221\textcolor{red}{35307}\\
4&-0.2164\textcolor{red}{9549}&-0.2210\textcolor{red}{9101}\\
5&-0.21646\textcolor{red}{238}&-0.22107\textcolor{red}{111}\\
6&-0.216469\textcolor{red}{51}&-0.221074\textcolor{red}{23}\\
7&-0.2164697\textcolor{red}{6}&-0.2210743\textcolor{red}{9}\\
8&-0.21646972&-0.22107437\\
9&-0.21646972&-0.22107437\\
10&-0.21646972&-0.22107437\\
11&-0.21646972&-0.22107437\\
12&-0.21646972&-0.22107437\\

\hline
\end{tabular}
\caption{\raggedright \small Projected and expectation value UCCSD correlation energy for STO-3G N$_2$ at $r=1.3\r{A}$, as a function of truncation order. The results are converged to within $10^{-8}\mathrm{E_h}$ by $o=8$.}
\label{tab:conv}
\end{small}
\end{table}

\begin{figure}[h]
\subfloat{\includegraphics[width=0.5\textwidth, center]{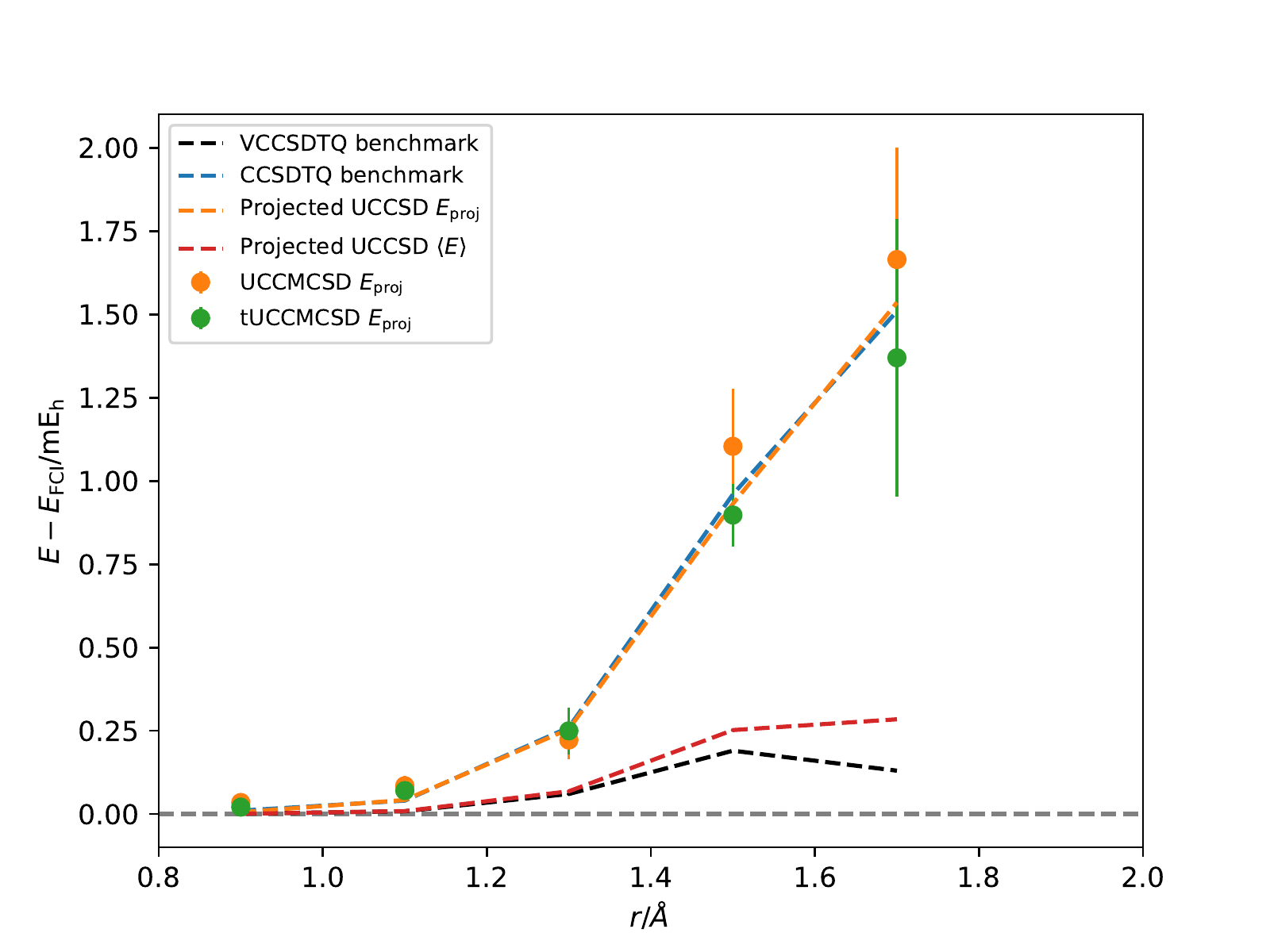}}
\caption{\raggedright \small Error in stochastic CCSDTQ and UCCSDTQ energy estimates relative to the FCI energy for N$_2$ in a STO-3G basis set. CCSDTQ and VCCSDTQ benchmark values are from Ref. \onlinecite{Cooper2010}. }
\label{fig:n2-uccsdtq}
\end{figure}

\textcolor{black}{Finally, we note that the stochastic approach we have implemented can be directly applied to higher orders of coupled cluster, with no additional complexity. For example, \Cref{fig:n2-uccsdtq} shows projected energy results for stochastic methods including up to fourth order excitations. These are significantly more accurate than their CCSD counterparts, but once again we observe that variational energy would provide a better quality estimator than the projected energy. It is worth noting that, due to the linear scaling of cluster selection with number of excitors in tUCCMC, we observe a slowing down of the trotterized method relative to full UCCMC, for identical calculation parameters.}

\section{Conclusions}

In this paper, we have developed a projective approach to the unitary coupled cluster method, based on solving the residual equations for a polynomially truncated unitary exponential ansatz or its Trotter approximation. We have further implemented a stochastic version of this method, within the framework of CCMC.

For two-electron systems, the UCCMCSD method shows good agreement with FCI, as expected. For larger systems, we find that the stochastic and deterministic results computed at the same polynomial truncation level agree, implying that no bias is introduced by our selection schemes. Further, we have shown that for N$_2$, the results quickly converge with polynomial truncation level, guaranteeing that this truncation does not introduce any meaningful error in the results. Finally, we have observed that, in the context of UCCSD, the expectation value of the energy appears to provide a better estimator for the correlation energy than the projected energy, approaching the value obtained by variational UCCSD, without requiring the explicit variational optimisation of the wavefunction with respect to the cluster coefficients, which leads to significantly more involved equations. Both computing this quantity and obtaining an unbiased estimator for it is more expensive than the corresponding procedure for the projected energy, so it is satisfying to note that while less accurate, the projected energy we obtain is comparable to CCSD and we therefore expect it to improve in a similar way with increased cluster orders. \textcolor{black}{Furthermore, unlike its conventional counterpart, the stochastic UCCMC approach and its trotterized approximation naturally extends beyond singles and doubles, allowing one to access higher accuracy methods in a unitary fashion. As such, we are optimistic that this approach could become a viable alternative to traditional coupled cluster, when high accuracy is required and a unitary \textit{ansatz} is preferable. UCCMC is expected to scale well with increasing cluster truncation levels, provided the polynomial truncation level is preserved. However, for very high cluster truncations, the error due to the finite polynomial order used may become higher and this would need to be increased appropriately. While tUCCMC suffers from no such errors, the current sampling algorithm used scales linearly with the number of excitors in the expansion and we therefore expect this step to become limiting in large enough Hilbert spaces.}

Given the renewed interest in unitary coupled cluster as a functional form in the quantum computing community, we believe that our stochastic method may be of interest as a means to provide a better--than--Hartree--Fock initial guess for the wavefunction or to screen the cluster amplitudes, as CCMC has been used before for conventional algorithms,\cite{Deustua2017, Deustua2018} helping to streamline the quantum algorithm, \textcolor{black}{by decreasing the complexity of the circuits required in a system-dependent, physically justified way}. These ideas will be explored in a further publication.

\section{Supplementary Material}
See supplementary material for details on the convergence of UCCMC and tUCCMC, as well as cluster coefficients for the LiH and N$_2$ systems.
\section{Acknowledgements}
M-A.F. is grateful to the Cambridge Trust and Corpus Christi College for a
studentship and A.J.W.T. to the Royal Society for a University Research 
Fellowship under Grant No. UF160398. Both thank Prof. Peter Knowles, Dr. David Mu\~noz Ramo and Nathan Fitzpatrick for useful discussions.
\section{Data Availability}
The data that support the findings of this study are openly available in the Apollo - University of Cambridge Repository at
https://doi.org/10.17863/CAM.60037.
\bibliography{./biblio}

\end{document}